\documentclass[useAMS,usenatbib]{mn2e}

\usepackage{epsfig}
\usepackage{natbib}
\usepackage{lscape}

\newcommand{\kms}{km~s$^{-1}$}

\title{Cosmic Flows: Univ. Hawaii 2.2m I-band Photometry}

\author[Courtois et al.]
{H\'el\`ene M. Courtois$^{1,2}$\thanks{E-mail: \texttt{h.courtois@ipnl.in2p3.fr}}, 
R. Brent Tully$^{2}$, 
Philippe H\'eraudeau $^{3}$\\ 
$^{1}$Universit\'e Lyon 1, CNRS/IN2P3/INSU, Institut de Physique Nucl\'eaire, Lyon, France\\
$^{2}$Institute for Astronomy, University of Hawaii, 2680 Woodlawn Drive, Honolulu, HI 96822, USA\\
$^{3}$Argelander-Institut f\"ur Astronomie, Auf dem H\"ugel 71, D-53121, Bonn, Germany}

\begin{document} 
\date{Accepted 2011 March 25th. Received xxxx; in original form 2011 March 4}

\pagerange{\pageref{firstpage}--\pageref{lastpage}} \pubyear{2011}

\maketitle

\label{firstpage}

\begin{abstract}
Within the "Cosmic Flows" project, $I$-band photometry of 524 relatively nearby galaxies has been carried out over the course of several years with the University of Hawaii 2.2m telescope and a camera with a 7.5 arcmin field of view.  The primary scientific purpose was to provide global magnitudes and inclinations for galaxies for the purpose of measuring distances through the correlation between galaxy luminosities and rotation rates.  The $1 \sigma$ accuracy on a total magnitude is 0.08 mag. The  observations typically extend to 7--8 exponential disk scalelengths so the data are useful for studies of the structural properties of galaxies. 
\end{abstract}


\section{Introduction}

Within the "Cosmic Flows" project, the motivation for this optical photometry program is the measurement of extragalactic distances based on the correlation between the rotation rates of galaxies and their luminosities \citep{1977A&A....54..661T}.  Other papers in this series discuss the compilation of rotation rate information \citep{2009AJ....138.1938C, 2011arXiv1101.3802C} and the synthesis of data into the final product of distances.  The global program involves many components which are being assembled for dissemination in EDD, the Extragalactic Distance Database, on the web at http://edd.ifa.hawaii.edu \citep{2009AJ....138..323T}.

The determination of a distance with the correlation between luminosities and linewidths requires knowledge of three observational parameters: measures of the rotation rate of a galaxy, the total luminosity of the system, and its inclination.  The latter two parameters are determined through surface photometry. 
 It must be asked if there are preferred passbands that optimize the correlation for the purpose of distance measurements.  The best correlation is expected to arise in a wavelength interval that samples the peak of the black body emission from old stars since this population should be the best thermal representative and contributor to the potential well.   
This consideration favors wavelengths in the $1-2~\mu$m range, where there is the added advantage of diminished obscuration compared with optical bands \citep{1979ApJ...229....1A}.

Unfortunately, ground-based observations at wavelengths longer than 0.8 microns are hampered by severe and variable sky emission from OH auroral lines, and beyond 2 microns by thermal emission.  As a consequence, imaging at these longer wavelengths is much more difficult, requiring constant monitoring of the sky and stacking of short exposures, and with a given integration one typically reaches one exponential scalelength less deep compared with what can be achieved at shorter wavelengths \citep{1996AJ....112.2471T}. From experience, it is found that the tightest correlations between luminosity and linewidth are found at $R$ and $I$ bands, at wavelengths between 6000~\AA\ and 8000~\AA\  \citep{2008ApJ...676..184T}.  Scatter is slightly higher at the redward bands $H$ and $K^{\prime}$ because of sky noise and higher at the shorter wavelength of $B$ band because of active star formation and obscuration.  The focus of the present discussion is on data collected in the sweet spot at Cousins $I$ band \citep{1979PASP...91..589B}.  We retain an ongoing interest in photometry at longer wavelengths  \citep{2009MNRAS.394.2022M} and will discuss observations at $K^{\prime}$ band relevant to our program in another paper.  

\section{Observations}

Our galaxy photometry program is a continuation of work discussed by \citet{1988ApJ...330..579P,1992ApJ...387...47P},  \citet{1996AJ....112.2471T}, and \citet{2000ApJ...533..744T}.  There is a discussion of the "Cosmic Flows" samples that are being collected by \citet{2011arXiv1101.3802C}.  The new photometry pertains to the samples identified in that reference as `calibrators' (galaxies in 13 clusters that define the slope of the rotation rate $-$ luminosity correlation and galaxies with independently established distances that set the zero point of the correlation), `SNIa' (galaxies that have hosted supernovae of type Ia), `V3K'(galaxies within 3000~\kms\ brighter than $M_{Ks} = -21$), and `PSCz' (galaxies within 6000~\kms\ selected by far infrared flux properties).

The material discussed here has been obtained since 2000, mostly since 2006, at the University of Hawaii 2.24m telescope using a Tek 2048 CCD.  The instrument is mounted at the f/10 Cassegrain focus and provides a 7.5 arcmin diameter field with a pixel scale of 0.22 arcsec.  Monitoring of the sky to assure photometric conditions has mostly relied on SkyProbe at the Canada-France-Hawaii Telescope\footnote{http://www.cfht.hawaii.edu/Instruments/Elixir/skyprobe}.  Only data acquired in photometric conditions are accepted. Table~\ref{tbl:log} provides a brief summary of the observing runs. 
A total of 651 images were obtained during 18.8 photometric of 30 scheduled nights, with some repetition, producing useful data for a final list of 524 individual galaxies. 

\begin{table}
\caption{Summary Observing Log}
\begin{tabular}{lcrl}
\hline\hline
Date & Phot./Nights & Gal. & Observers\\
\hline
2000/02/01 & 5.0 of 5 & 107 & BT \\
2006/12/19 & 5.5 of 6 & 153 & KC, HC, LR, BT \\
2007/03/12 & 2.5 of 8 &  172 & KC, HC, LR, BT, MZ \\
2007/09/14 & 1.8 of 3 &   84  & BT, MZ \\
2007/10/06 & 1.0 of 3 &   48  & BT, MZ \\
2008/02/28 & 3.0 of 5 &  87 & HC, BT \\
\hline\hline 
\end{tabular}
\label{tbl:log}
\end{table}

\begin{figure} 
\begin{centering}
\includegraphics[scale=0.44]{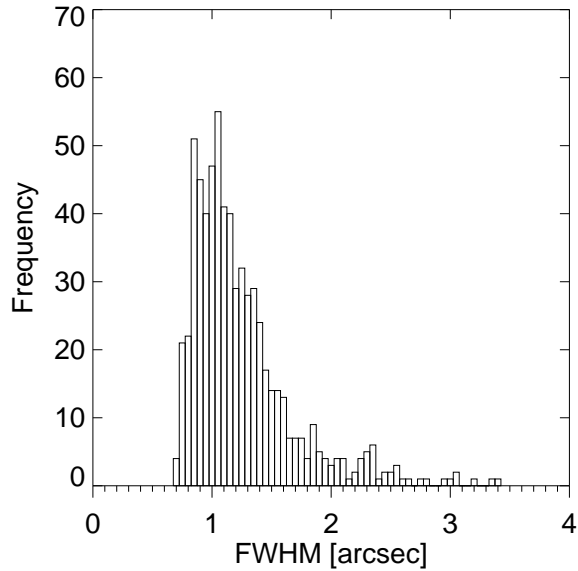}
\caption{Histogram of the FWHM for the 622 observations acquired at the University of Hawaii 2.2m telescope. } 
\label{seeing}
\end{centering}
\end{figure}

Most of the observations were made through a filter that approximated Cousins $I$ 
 with a standard integration of 300 sec.  Instances of nuclear saturation (conservatively, counts in excess of 30,000) prompted the acquisition of complementary short exposures.  For a small fraction of the galaxies, images were also obtained in Cousins $R$ (300 sec integrations) and Johnson $B$ (600 sec integrations).  The observations were usually but not always guided.  Image quality is erratic, mostly ranging from FWHM 0.7" to 1.5" with a median seeing of 1.2 (see Fig.~\ref{seeing}), but this was not a great concern since our interest is in global photometry and the target galaxies have typical dimensions of 1--5 arcmin.  
Photometric calibrations in the Vega system were made with observations of \cite{1992AJ....104..340L} standard fields on several occasions each night and over air masses ranging up to 2 and beyond.  Flat fields were built from multiple twilight exposures of blank sky supplemented by dome flats.  The properties of the mechanical shutter were studied in order to compensate for variations across the field caused by the finite time required to open and close (Zissell 2000).
The telescope is at a latitude of $+20^{\circ}$.  Targets were observed as far north as $+85^{\circ}$ and as far south as $-40^{\circ}$.  We report on observations of 524 galaxies.

\subsection{Flagging}
Out of 651 observations, a total of 34 images were rejected. The main reason was that longer exposures were available for the same galaxies, 2 galaxies (PGC 70163 and PGC 63545) have only 2 images with 30 sec exposures, all the others have 100s or 300s exposures. A few images were excluded because of problems with flat-fielding or telescope tracking.

\section{Astrometric calibration}
Flat-fielded images were astrometrically calibrated using the Scamp software Version 1.4.6 \citep{2006ASPC..351..112B}.
Source extraction was first performed with the SExtractor software Version 2.4.4 \citep{1996A&AS..117..393B} using a detection threshold equal to 1.8.  As reference for the astrometry,
Scamp was run using the USNO-B1 catalog \citep{2003AJ....125..984M}, which is complete down to $V=21$ and has an astrometric accuracy of 0.2 arcsec. Since astrometry is not our main concern and the field of view of our observations is relatively small (thus there are only a small number of stars available to derive an astrometric solution), a low order polynomial (DISTORT\_DEGREES=1) was used in Scamp. 
Images where then resampled with the Swarp software Version 2.17.1 \citep{2002ASPC..281..228B} to apply the astrometric solution to the images. The pixel size was kept the same; i.e. 0.22 arcsec/pixel.

\begin{figure} 
\begin{centering}
\includegraphics[scale=0.42]{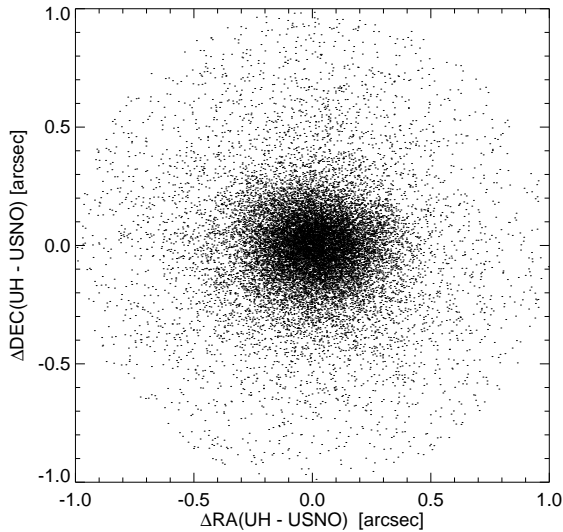}
\caption{Differences between the positions of 18748 stars detected on our images (UH) after applying the astrometric solution and those in the USNO-B1 catalogue (USNO) along R.A. and DEC. in arcsec. The matching between the 2 catalogues was performed with an accuracy of 1 arcsec. Only stars with an $R$-band magnitude brighter than 19 in the USNO are shown.}
\label{astro_usno}
\end{centering}
\end{figure}

Figure~\ref{astro_usno} shows the difference between the source position detected by SExtractor on the regridded images and the USNO-B1 catalogue for 18748 stars with an $R$-band magnitude brighter than 19 in the USNO catalog. A search radius of 1 arcsec was used.
The mean difference is smaller than 0.01 arcsec in R.A and DEC. with a dispersion of 0.27 arcsec and 0.26 arcsec, respectively.
All our 651 exposures were used for this comparison. 

\begin{figure} 
\begin{centering}
\includegraphics[scale=0.42]{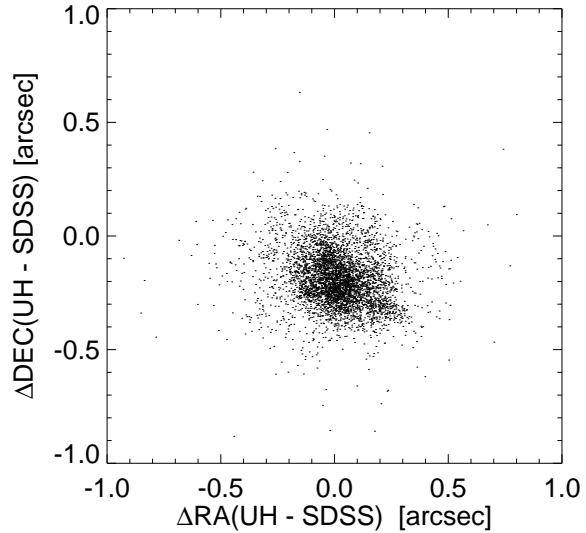}
\caption{Differences between the positions of 4720 stars detected on our images (UH) after applying the astrometric solution and those in the SDSS-DR7 catalogue (SDSS) along R.A. and DEC. in arcsec. The matching between the 2 catalogues was performed with an accuracy of 1 arcsec. Only stars with a SExtractor FLAG parameter equal to 0 were selected. Out of 622 images observed at UH, 301 have been observed by the SDSS.}
\end{centering}
\label{astro_sdss}
\end{figure}

Since the Sloan Digital Sky Survey (hereafter SDSS) catalog is used for searching for stars in common with our images for photometric calibration purposes (see Sect.~\ref{photo}), it is important to check that the astrometry is reliable enough for that purpose.
Fig.~\ref{astro_sdss} shows the comparison of 4720 stars (as defined by SDSS) in common between our images and the SDSS seventh data release (DR7) catalog \citep{2009ApJS..182..543A} within a search radius of 1 arcsec. Only unflagged stars (with a SExtractor FLAG parameter equal to 0) were selected. Out of 651 images, 301 have been observed by the SDSS.     
The small differences and offsets in the astrometry of the two surveys most probably come from the different methods adopted to determine the source positions and the astrometric solution, and the choice of different reference catalogues for the astrometric calibration of the SDSS which differ from the USNO-B1 we use in this work (see details at http://www.sdss.org/dr7/algorithms/astrometry.html for the SDSS).
The astrometry is overall very good with a median between our observations and the SDSS DR7 equal to 0.01 and -0.20 arcsec along R.A. and DEC with a dispersion of 0.15 and 0.13 arcsec, respectively. 

\section{Photometry}
\label{photo}
Initial reduction steps involved the construction of flat fields from summations of typically 5 twilight exposures.  For the conversion from counts to flux in magnitudes it was expected at the time of the observations that this transformation would exclusively be based on aperture photometry of \citet{1992AJ....104..340L} standards observed every night throughout a run.  These observations were analyzed with flux corrections applied to account for shutter effects with short exposure times.  
The shutter effects are negligible for the normal exposures on galaxy targets but slightly impact the Landolt standard star exposures.
In addition, for fields within the parts of the sky covered by the SDSS, counts were transformed to magnitude scales from calibrations provided by stars in each field with known SDSS psf-fit $r,i$ magnitudes.

\subsection{Photometric calibration based on the SDSS}
Although Landolt standards were regularly observed to calibrate the photometry, ultimately it was determined that greater accuracy was afforded by the use of stars within fields with magnitudes determined by the SDSS.  
Indeed, the SDSS provides many well-calibrated stars that can be used for calibration purposes, although the accuracy is not quite as good as with the Landolt standard stars (typically better than 0.02 mag for stars brighter than 20 in the SDSS $r$ filter, compared to one part in a thousand for Landolt standard stars).

About half of our observations (301 out of 651) are within the SDSS.  In these cases an exposure can be calibrated differentially independent of monitoring of the photometric conditions through the night which is a major advantage. 
In practise, SExtractor was run on the astrometrically calibrated images and sources were matched to the SDSS star catalogue with a search radius of 1 arcsec to derive the photometric zeropoints.
The magnitudes of the stars for the observations covered by the SDSS were first retrieved from the SDSS DR7 catalog as the $PSFMAG$ estimate and color-corrected to the Cousins photometric system at $I$ band. 
Two sets of transformation equations were tested for that purpose. In the end, the transformation of \citet{2002AJ....123.2121S} was preferred over the ones of \citet{2005AJ....130..873J} because it provides smaller dispersions on the zero points averaged over the same time intervals; i.e. over one night or a couple of nights as defined in Table~\ref{tab: median_zp}.
Typically, the zero point for each of the 301 observations with SDSS data was derived from the mean of 12 stars with a 2-sigma rejection and has an accuracy of 3 percent. 
Fields without SDSS observations or with less than 5 SDSS stars were calibrated with the median of the zero points grouped per night or for several nights as given in Table~\ref{tab: median_zp}; they have standard deviations between 0.01 and 0.03 mag.
A mean value of 0.043 mag/airmass was used to account for atmospheric extinction in the fields without SDSS coverage.

We have compared the SDSS-based calibration with a calibration obtained with Landolt standard stars observed during our runs and we compared results with the separate calibrations to the external data set discussed in the next section.  The SDSS and Landolt calibrations agree in zero point within the statistical uncertainties.  In the comparisons with the external data set, aside from a zero point issue that will be discussed next, the SDSS-based calibration gives the tighter correlations.  Hence, the SDSS-based calibration is preferred to the Landolt standard star calibration. 

\begin{table}
\begin{minipage}[t]{\columnwidth}
  \caption{The color correction transformations from the Sloan
    $g^{\prime},r^{\prime},i^{\prime}$ filters to the Johnson-Cousins $B$, $V$, $R_c$, $I_c$ 
    filters. Taken from Table 7 of \citet{2002AJ....123.2121S}.}
\label{tab: color_correction}
\centering
\begin{tabular}{c | r | l}
\hline\hline

\# & Magnitude/Color & Transformation \\
\hline
(1) & $B$ & $= g^{\prime} + 0.47\cdot(g^{\prime}-r^{\prime}) + 0.17$\\
(2) & $V$ & $= g^{\prime} + 0.55\cdot(g^{\prime}-r^{\prime}) + 0.03$\\
(3) & $(V-R_c)$ & $= 0.59\cdot(g^{\prime}-r^{\prime}) + 0.11$\\
(4) & $(R_c-I_c)\quad \mathrm{for}\;(r^{\prime}-i^{\prime}) < 0.95$ & $= 1.00\cdot\mathrm{(r^{\prime}-i^{\prime})}+0.21$\\
(5) & $(R_c-I_c)\quad \mathrm{for}\;(r^{\prime}-i^{\prime}) \ge 0.95$ & $= 0.70\cdot\mathrm{(r^{\prime}-i^{\prime})}+0.49$\\
\hline
\hline
\end{tabular}
\end{minipage}
\end{table}

\begin{table}
\begin{minipage}[t]{\columnwidth}
  \caption{The mean zero points derived from SDSS photometry. Columns give the Julian dates of the beginning and end of the period, the median (ZP) and standard deviation (EZP) of the zero points and the number of measurements (N).}
\label{tab: median_zp}
\centering
\begin{tabular}{c | c | c | c| c}
\hline\hline
Begin       & End        & ZP  & E\_ZP & N \\
Julian Day  & Julian Day & MAG & MAG &  \\
\hline 
  2451079.00   & 2452905.00 &  24.244  &  0.021   & 65  \\ 
  2452905.00   & 2454134.00 &  24.269  &  0.012   & 27  \\ 
  2454134.00   & 2454263.00 &  24.237  &  0.020   & 82  \\  
  2454263.00   & 2454373.00 &  24.275  &  0.026   & 13  \\ 
  2454373.00   & 2454432.00 &  24.301  &  0.012   & 12  \\ 
  2454432.00   & 2454850.00 &  24.289  &  0.027   & 61  \\ 
\hline
\hline
\end{tabular}
\end{minipage}
\end{table}

\subsection{The UH 2.2m Photometric System}

\begin{figure} 
\includegraphics[scale=0.39]{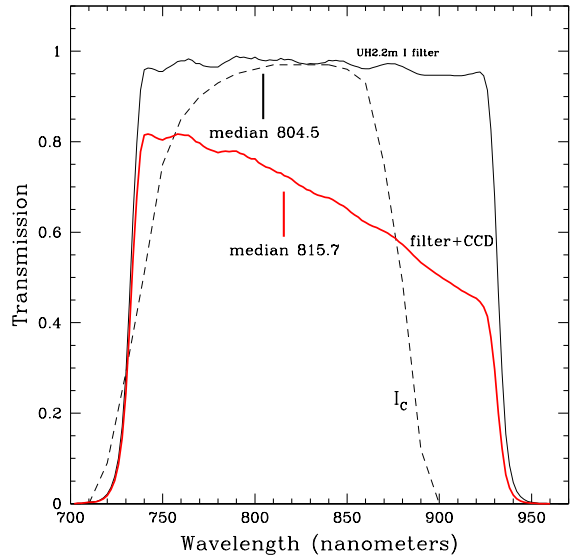}
\caption{
Comparison of filter transmission characteristics.  The Cousins filter passband $I_c$ is illustrated by the dashed curve with median transmission response at 804.5 nm.  The $I$ filter used on the University of Hawaii 2.2m telescope has the transmission properties shown by the thin solid curve. The passband convolved with the response of the Tektronix 2048 CCD detector is shown by the heavy (red) solid curve with median transmission response at 815.7 nm.  The dashed curve representing $I_c$ and the heavy solid curve representing the Hawaii system are normalized to the same area under each curve.}
\label{filt}
\end{figure}

Although the observations with the University of Hawaii 2.2m telescope approximate the Cousins $I$ system, $I_c$, in detail there is a systematic offset between the magnitudes of the galaxies derived from our observations and the $I_c$ magnitudes found in the literature. Figure~\ref{filt} illustrates the relative transmissions of the Cousins passband with a photoelectric detector and the filter plus Tektronics 2048 CCD detector used with this experiment.
 The UH 2.2m passband transmits radiation 50 nm to the red of the $I_c$ cutoff and this results in a systematic difference between the total magnitudes of galaxies derived by the procedures described in Sect.~\ref{archangel} and $I_c$ magnitudes given in the literature.
 
\begin{figure}
\begin{tabular}{l}
\includegraphics[width=0.5\textwidth]{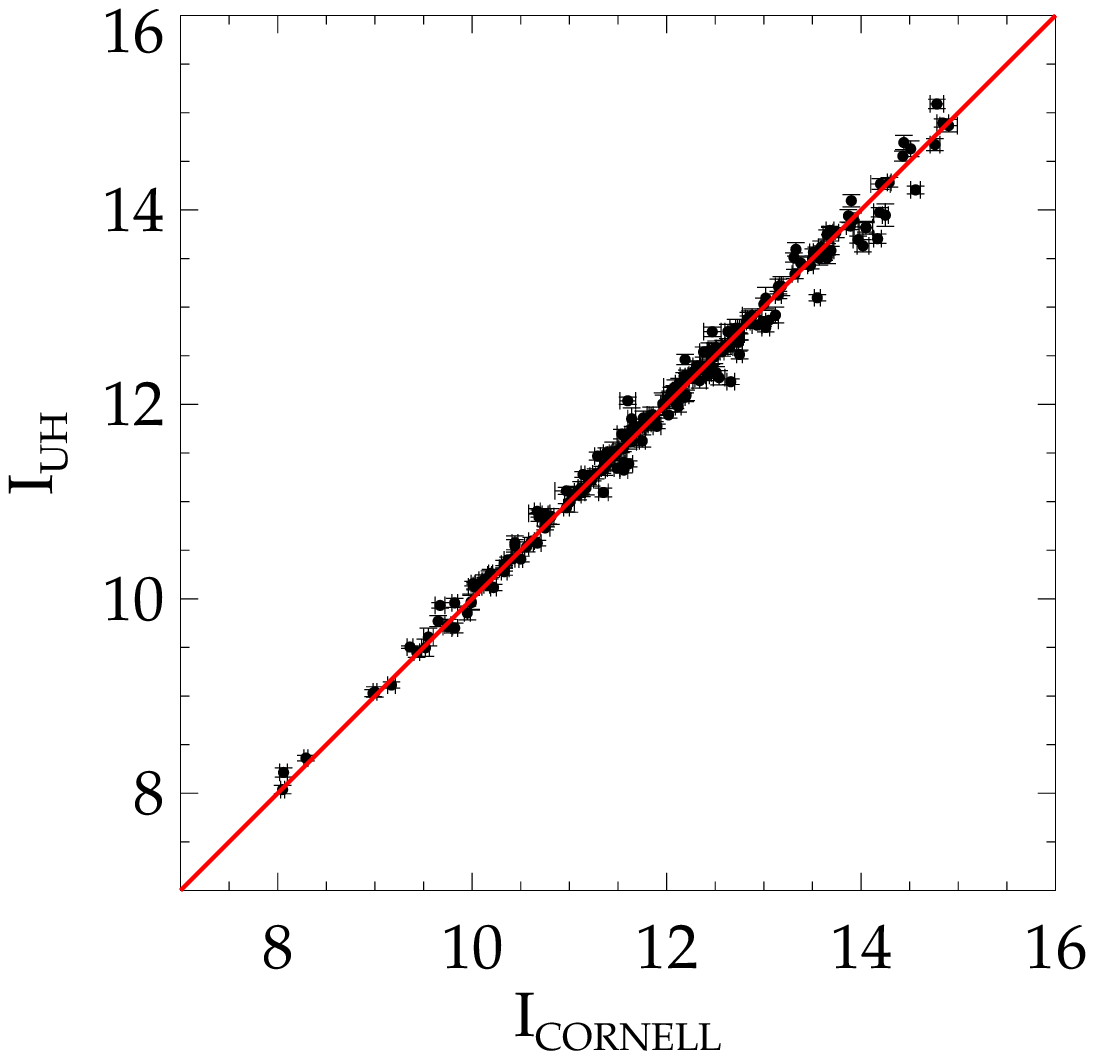}\\
\includegraphics[width=0.5\textwidth]{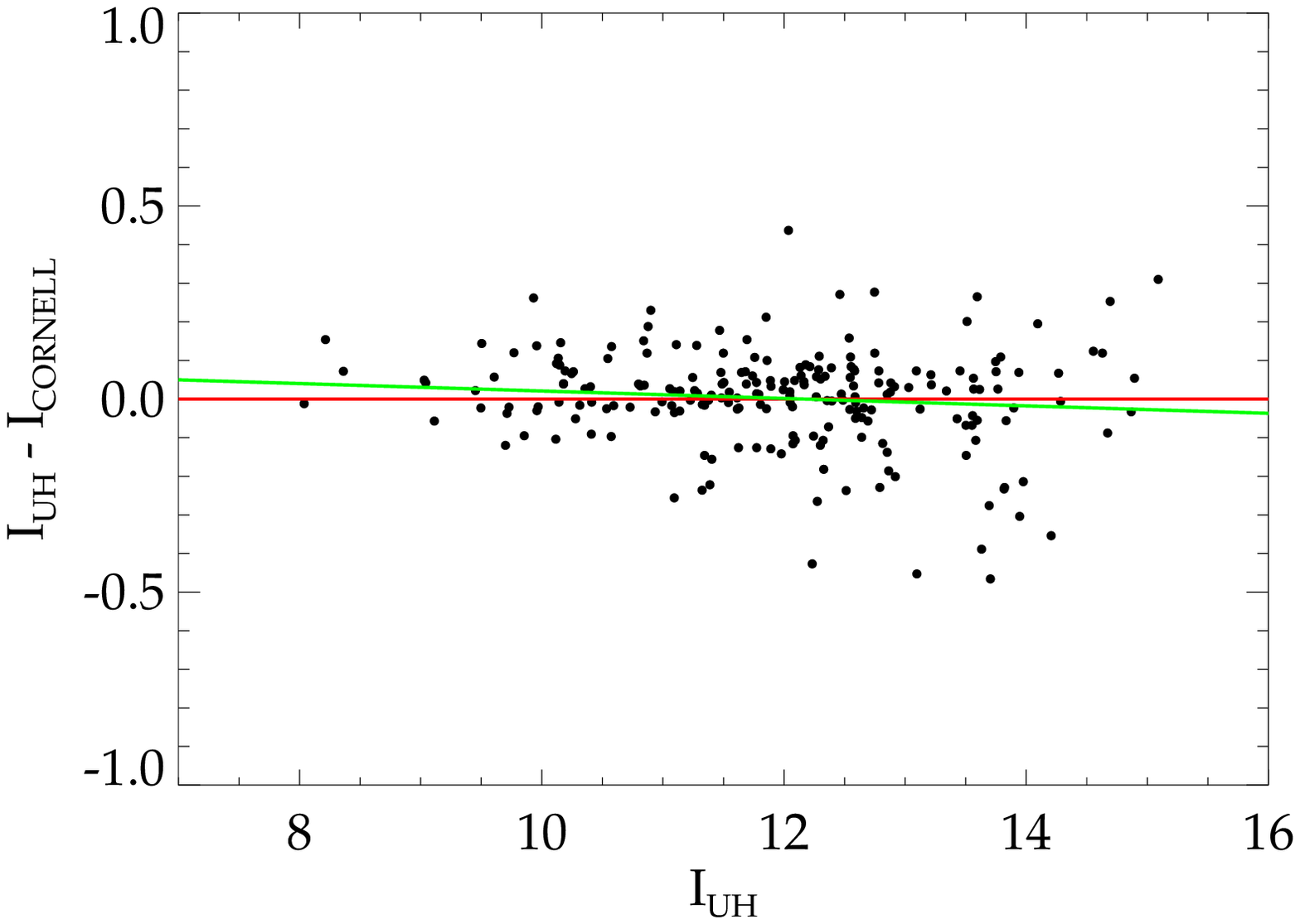}
\end{tabular}
\caption{Comparison of the total magnitudes obtained with Archangel and calibrated with SDSS stars with those of \citet{2007ApJS..172..599S} after applying a systematic offset of 0.12 mag to account for the difference between the University of Hawaii filter transmission and the Cousins filter. Mutliple observations of the same galaxies were combined before the comparison.  A differential comparison is shown in the bottom panel.  The solid red line shows the 1:1 correspondence. The slanted green line in the bottom panel illustrates a best fit but the departure from 1:1 correspondence is not significant.}
\label{Cornell}
\end{figure}

The largest homogenous compilation of photometry in the $I_c$ passband presently available has been published by the Cornell group  \citep{2007ApJS..172..599S}.
There are 286 galaxies from this source in common with our sample.   A systematic offset of 0.12 mag was found that is attributed to the extended red transmission of the UH 2.2m passband (Hawaii brighter).
A comparison of final $I_{uh}$ magnitudes and \citet{2007ApJS..172..599S} measurements is shown in Fig.~\ref{Cornell}.  The Hawaii values are the weighted mean of the total magnitudes in the cases of multiple observations of the same galaxy and have been adjusted faintward by 0.13 mag.   
The slope of the correlation is unity within the errors over a 6 magnitude range, with a dispersion of 0.11 mag in $<I_{UH} - I_{Cornell}>$.  If the Hawaii and Cornell errors are comparable then $1 \sigma$ errors of 0.08 mag are implied for each source.  A small number of mildly discrepant measurements are noted in Table~\ref{tbl:discrep}.

\begin{table}
\caption{Discrepant measurements with Cornell data}
\begin{tabular}{lccccr}
\hline\hline
PGC &      $I_C$ & $err_C$ & $I_{UH}$ & $err_{UH}$ & $I_{UH} - I_C$  \\
\hline
   7706 & 14.02 &  0.06 & 13.63 &  0.06 &  -0.39 \\
   8232 & 11.60 &  0.08 & 12.04 &  0.04 &  +0.44 \\
   9816 & 14.56 &  0.05 & 14.21 &  0.04 &  -0.35 \\
   9895 & 13.55 &  0.03 & 13.10 &  0.03 &  -0.45 \\
  19558 & 14.25 & 0.03 & 13.95 & 0.12 &  -0.30 \\
  24870 & 14.78 & 0.07 & 15.09 & 0.05 &  +0.31\\
  26455 & 12.66 & 0.04 & 12.23 & 0.03 & -0.43\\
  50728 & 14.17 & 0.04 & 13.70 & 0.05 & -0.47\\
\hline\hline 
\end{tabular}
\label{tbl:discrep}
\end{table}

\subsection{Archangel Photometry}
\label{archangel}

\begin{figure*}
\vspace{10mm}
\begin{tabular}{ll} 
\includegraphics[width=0.7\textwidth]{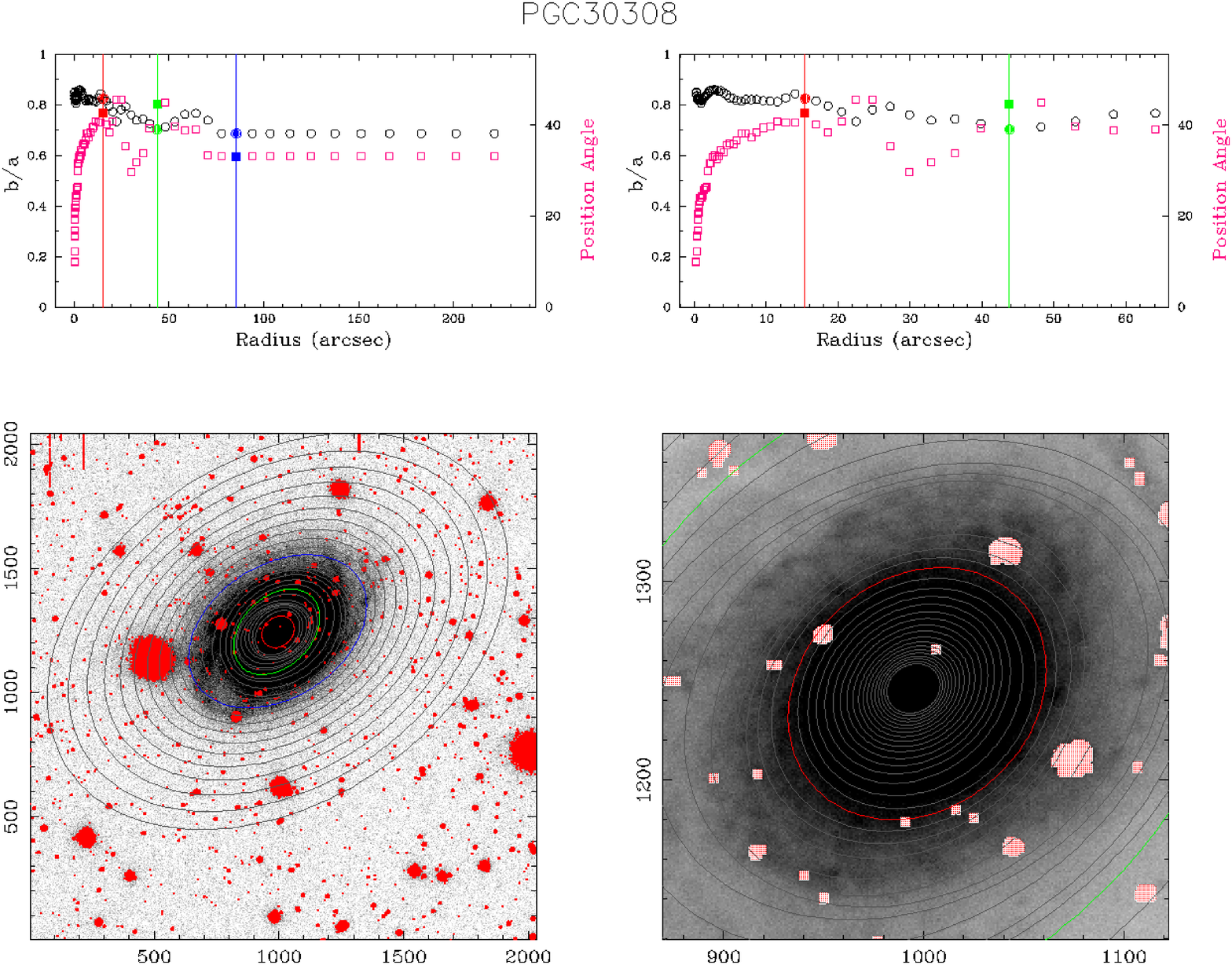}\\
\includegraphics[width=0.4\textwidth]{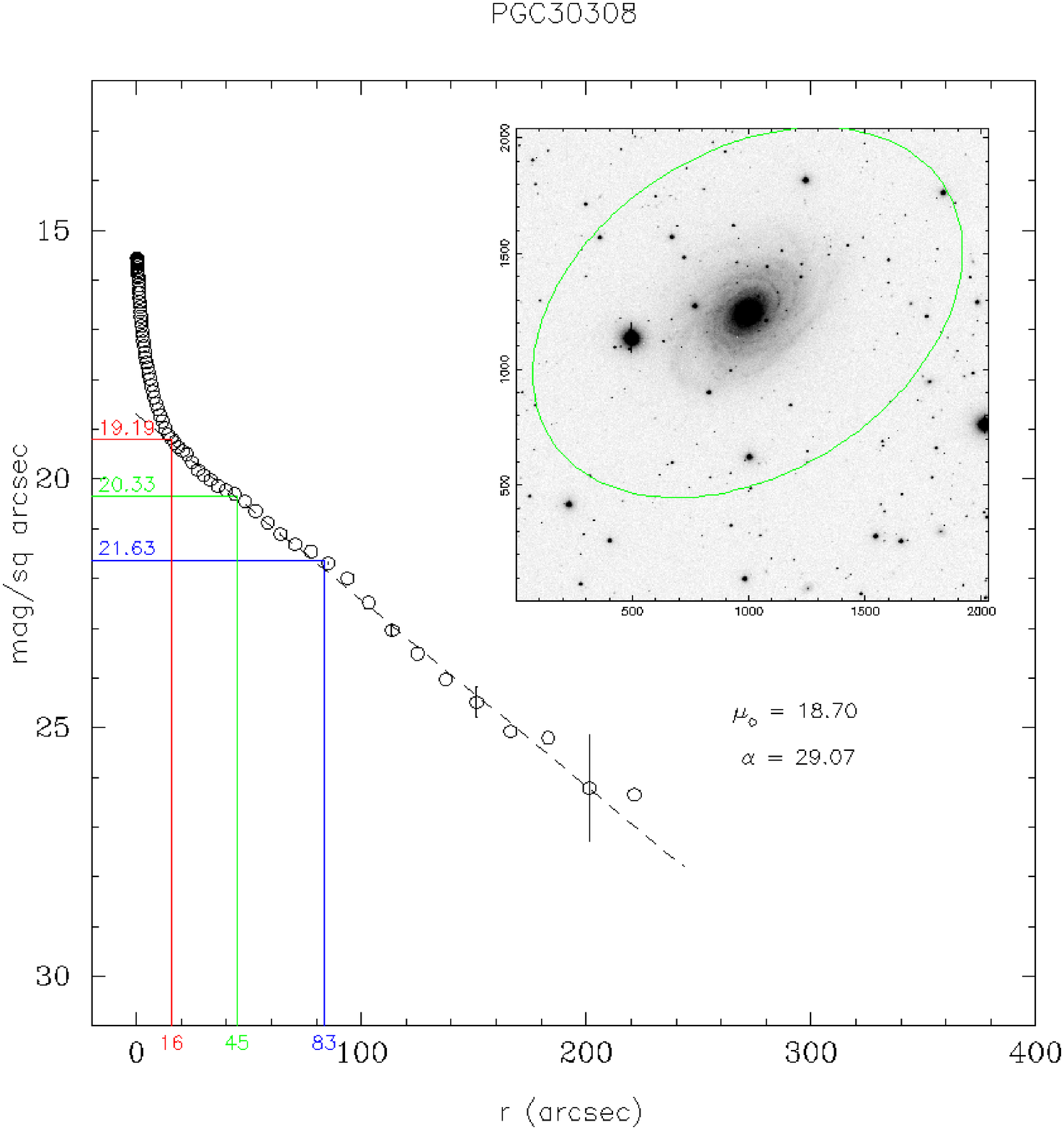} &
\includegraphics[width=0.4\textwidth]{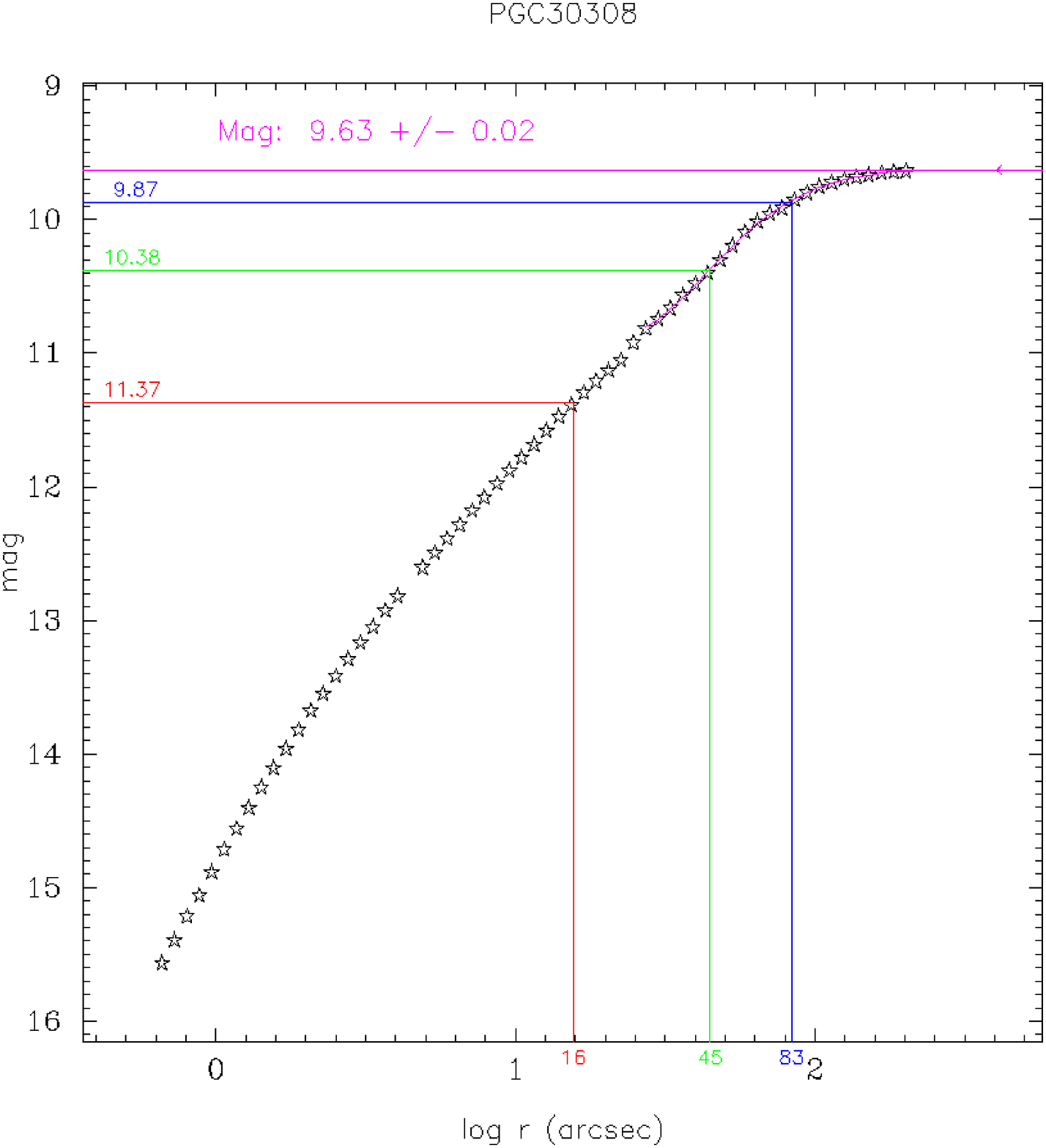}\\
\end{tabular}
\caption{Photometry of NGC 3223.  Graphic output from the Archangel photometry program. {\it Top:} in black the run with radius of ellipticity measures, b/a = ratio of minor to major diameter,  and in red the run with radius of the position angle of the major axis.  The radii enclosing 20, 50, and 80 percent of the total light of the  galaxy are indicated by the red, green, and blue vertical strips respectively.  The portion of the left panel at small radii is seen more easily with the expanded scale of the right panel.  {\it Middle:} I band CCD image of the target with stars and bad pixels masked and ellipse fits shown at radial increments.  The red, green, and blue ellipses enclose 20, 50, and 80 percent of the total light.  {\it Bottom:}  the run of surface brightness with radius is shown on the left and the growth of enclosed magnitude with radius is shown on the right.  Radii enclosing 20, 50, and 80 percent of the light are indicated in red, green, and blue respectively.  The surface brightness and enclosed magnitude at each of these radii are identified.  An I band image of the galaxy is shown in the inset.  The green ellipse is drawn at the surface brightness level 27.0 mag./sq. arcsec.}
\label{photometry}
\end{figure*}

The flattened and flux calibrated images were analyzed with the Archangel software package developed by \citet{2007astro.ph..3646S}\footnote{http://abyss.uoregon.edu/$\sim$js/archangel}.  The suite of programs performs such procedures as masking of stars and flaws, ellipse fitting at expanding radii from the galaxy center, and compression of two-dimensional information into one-dimensional growth curves of surface brightness as a function of radius and total luminosity as a function of radius.  An example of output from the Archangel analysis is shown in Figure~\ref{photometry}.

\subsection{Magnitudes and Inclinations}

The two parameters we care about most for the determination of distances through the correlation between luminosities and rotation rates are total magnitudes and inclinations.  Consider, first, issues that affect the measurement of luminosities.  A significant source of uncertainty arises from the setting of the sky level, where a change of 1 count per pixel typically affects total fluxes by 2\% (0.02 mag).  Almost all our targets are modest in size compared with the CCD field so there is reasonable control of the sky level.  If sky is set properly then the magnitude growth curve should go asymptotically flat at large radii.  One is suspicious of a poor sky setting if surface brightness as a function of radius either flares or drops precipitously at the sky level, although the latter occurrence is not physically excluded \citep{2003ApJ...582..689M}.

It is never possible to directly detect 100\% of the light of a galaxy.  Measurement is made to an isophotal level dictated by the telescope, exposure time, and the brightness of the sky.  The interest of this program is with spiral galaxies that characteristically decay exponentially in luminosity with radius.  The luminosity, $L$,  of a galaxy grow as 
\begin{equation}
L_x \propto \int_0^x x e^{-x/\alpha} {\rm d}x.
\end{equation}
 The radial dependence is described by the scalelength, $\alpha$.  Assuming the exponential decay in light can be extrapolated, the contribution lost below the sky level can be estimated and added to what is observed to give a `total magnitude'  \citep{1996AJ....112.2471T}.  Fortunately the extrapolation from an isophotal to a total magnitude is almost always small in our cases because our exposures typically capture 7--8 exponential scalelengths.  Extrapolations are usually below 0.02 mag and uncertainties are less than half the extrapolation.  For the small extrapolations that are required we use the procedure involving rational function fits to the magnitude growth curve at large radii provided within Archangel.  The only situations where the extrapolations are significant are either with galaxies that extend beyond the CCD field or with extremely low surface brightness galaxies.

The other product of importance to us provided by the photometry program is inclinations, $i$, derived from a measurement of ellipticities.  By convention, a face-on galaxy has $i=0$.  The inclination is derived from the observed ratio of the minor and major axes, $b/a$, under the assumption that a galaxy is a prolate ellipsoid with a thickness $b/a = q_0$.  The inclination is then given by the formula
\begin{equation}
{\rm cos}~i = \sqrt{{(b/a)^2 - q_0^2}\over{1 - q_0^2}}
\end{equation}
The question of the optimal value of $q_0$ will not be debated here.  The {\it Hawaii Photometry} catalog records only the observable $<b/a>$, determined as described below. 

The orientation of a galaxy must be known to de-project velocities of the rotating disk, with corrections increasing toward face-on, and to account for the effects of obscuration, which increase toward edge-on.  Perhaps surprisingly, the latter is not a severe problem.  Corrections for obscuration are substantial but predictable  \citep{1998AJ....115.2264T}.  After applying standard recipes, the scatter in luminosity--linewidth plots is not significantly greater for the most edge-on galaxies ($i \sim 90\degr$).  The greater concern with the measurement of inclinations is with the rectification of velocities.  The problem is minor for inclinations $i > 60\degr$ since the de-projection $W^i = W / {\rm sin}~i$ is small.  However the de-projection adjustment explodes as $i \rightarrow 0\degr$.  Inclination uncertainty dominates the scatter in luminosity--linewidth plots for galaxies with $i < 45\degr$.

If ultimately for the purpose of distance determinations we adopt the policy of excluding galaxies more face-on than 45\degr\ then the problematic interval for inclination measurements is $45\degr - 60\degr$.  From experience, spiral galaxies separate roughly equally into three classes.  In about a third of cases, ellipticities and major axis position angles are roughly constant across a wide range of radii and the inclinations of these galaxies can be considered reliable at the level of $2\degr - 3\degr$.  In another third of cases, uncertainties degrade to $3\degr - 5\degr$.  Unfortunately, in roughly the last third of cases, ellipticities and position angles vary by large amounts as a function of radius.  The reasons for these oscillations might be a large bulge or a bar or prominent spiral structure or a warp.  The ellipticity and position angle measurements can depend on the specifics of the ways these features are projected.  Anomalies can occur on all scales. In difficult cases, inclinations can be wrong by 5\degr\ to 10\degr, and inclination errors can be the dominant source of luminosity--linewidth scatter.

From experience, it is found that a decent measure of ellipticity can be given by averaging over values obtained for fitted ellipses between the radii enclosing 50\% and 80\% of the total light of a galaxy.
The distortions to ellipticity measurements from bulges and bars are frequently severe inside the radius containing 50\% of the light.  Beyond the radius enclosing 80\% of the light the signal rapidly becomes too weak for a reliable measure of axial ratio and we freeze the ratio at the value of the outermost reliable measurement.   Hence, our recorded estimate of the ratio of the minor to the major axis, $<b/a>$, is the average of all measures $(b/a)_i$ between the radius enclosing 50\% of the light, $a_e$, and the radius enclosing 80\% of the light, $a_{80}$.  The uncertainty that is tabulated is the r.m.s. dispersion in the measures contributing to the average.

All results obtained through the automated process that has just been described have been scrutinized by eye and, in instances, results have been modified and a large error has been assigned.    The estimation of inclinations from the assumption that galaxies are circular disks tilted from the line-of-sight is an uncertain business.  In spite of these travails our measured inclinations pass a necessary test: there are no systematic deviations from the mean luminosity--linewidth correlation as a function of inclination. 

\section{Summary}

Summary data and graphic displays for all the galaxies that have been observed for the "Cosmic Flows" project (http://ifa.hawaii.edu/cosmicflows) are found in EDD in the catalog {\it Hawaii Photometry}.  The parameters provided by Archangel from multiple observations of a single galaxy were combined as the weighted mean using the uncertainty on the total magnitude as weight. 
Table~\ref{final_results} provides an example of few lines of the final values from the photometry of the 524 galaxies. The full table is available digitally with this paper. The graphic material is accessed by selecting on the common name of a candidate and includes displays of star masks and ellipse fits, major axis position angle and ellipticity as functions of radius, surface brightness profiles, and magnitude--radius growth curves.  The radial information used for the construction of the plots is provided in ascii tables that can be downloaded. The tabulated data on the database includes information about the observations (detector, date, filter, etc.) and the measured parameters given in Table~\ref{final_results}.  By column in this table:

\medskip
\noindent
1. PGC (Principal Galaxies Catalogue) name.

\noindent
2. Common name.

\noindent
3. Total exposure in seconds.

\noindent
4. Number of exposures.

\noindent
5. $a_{27}$: Radius at 27 mag/arcsec$^2$ in arcsec.

\noindent
6. $m_{27}$: Magnitude within 27 mag/arcsec$^2$ isophote.

\noindent
7. $m_{tot}$: Total asymptotic magnitude.

\noindent
8. $e_m$: Uncertainty in total magnitude.

\noindent
9. $SB_0$: Extrapolated exponential disk central surface brightness.

\noindent
10. $\alpha$: Exponential disk scale length in arcesc.

\noindent
11. $b/a$: Ratio of minor to major axis dimensions.

\noindent
12. $e_{b/a}$: Error in ratio of minor to major axis dimensions.

\noindent
13. $PA$: Position angle of major axis, east of north.

\noindent
14. $a_{80}$: Major axis radius containing 80\% of light, arcsec.

\noindent
15. $SB_{80}$: Surface brightness at radius $a_{80}$.

\noindent
16. $a_{e}$: Major axis radius containing 50\% of light.; effective radius, arcsec.

\noindent
17. $SB_{e}$: Surface brightness at radius $a_{e}$.

\noindent
18. $avSB_e$: Average surface brightness within effective radius $a_e$.

\noindent
19. $a_{20}$: Major axis radius containing 20\% of light, arcsec.

\noindent
20. $SB_{20}$: Surface brightness at radius $a_{20}$.

\noindent
21. $avSB_{20}$: Average surface brightness within effective radius $a_{20}$.

\noindent
22. $C_{82}$: Concentration index $a_{80}/a_{20}$; ratio of radii containing 80\% and 20\% of light.

\section*{acknowledgements}
It was a delight to find that the photometry package Archangel developed by James Schombert and made available at http://abyss.uoregon.edu/$\sim$js/archangel is quite robust and user friendly.  L. Rizzi, K. Chiboucas, and students N. Bonhomme, M. Zavodny participated in some of the observing runs or have given help with early steps in the reductions.  This research made profitable use of the Sloan Digital Sky Survey and the VizieR catalogue access tool, CDS, Strasbourg, France.   RBT acknowledges support from the US National Science Foundation award AST-0908846.

\onecolumn
\begin{landscape}
\begin{table}
\caption{Parameters of the photometry}
\begin {tabular}{rlrrrrrrrrrrrrrrrrrrrr}
\hline\hline
  PGC  & Name & Exp & $N_e$ & $a_{27}$ & $m_{27}$ & $m_{tot}$ & $e_{m}$ &  $SB_0$ &  $\alpha$ &  $b/a$ &  $e_{b/a}$ &  $PA$ & $a_{80}$ & $SB_{80}$ & $a_e$ &  $SB_e$ & $avSB_e$ & $a_{20}$ & $SB_{20}$ & $avSB20$ &  $C_{82}$\\
\hline
    303 &  NGC7819   & 600 & 2 & 118 & 12.41 & 12.44 &  0.04 & 20.19 & 18.79 &  0.47 &  0.03 & 122 &  70 & 24.19 & 31 & 22.07 & 21.06 &  8 & 20.71 & 19.74 &   7.9\\
    608 &  NGC7836   & 300 & 1 &  64 & 12.32 & 12.31 &  0.06 & 19.14 &  8.90 &  0.53 &  0.03 &  44 &  17 & 21.30 &  8 & 18.90 & 18.03 &  2 & 17.85 & 17.04 &   7.0\\
    652 &  NGC0009   & 300 & 1 &  59 & 13.46 & 13.43 &  0.05 & 19.20 &  8.00 &  0.47 &  0.02 &  86 &  28 & 22.88 & 14 & 21.12 & 20.32 &  6 & 19.95 & 19.36 &   4.8\\
    963 &  UGC00139  & 300 & 1 &  98 & 12.86 & 12.82 &  0.03 & 20.01 & 15.00 &  0.44 &  0.04 & 170 &  45 & 23.32 & 21 & 21.47 & 20.43 &  8 & 20.06 & 19.53 &   5.6\\
   1160 &  NGC0063   & 600 & 2 &  80 & 11.17 & 11.24 &  0.07 & 17.57 &  9.30 &  0.55 &  0.05 &  11 &  29 & 20.61 & 17 & 19.54 & 18.87 &  8 & 18.68 & 18.16 &   3.9\\
   1288 &  PGC001288 & 300 & 1 &  67 & 12.89 & 12.90 &  0.05 & 18.64 &  8.70 &  0.46 &  0.03 &  58 &  29 & 22.01 & 18 & 20.98 & 20.23 &  7 & 20.00 & 19.45 &   4.1\\
   1592 &  UGC00243  & 300 & 1 & 137 & 11.85 & 11.90 &  0.06 & 18.52 & 17.40 &  0.22 &  0.01 &  95 &  63 & 22.66 & 33 & 20.74 & 19.85 & 19 & 19.73 & 19.14 &   3.4\\
   \hline\hline
\end{tabular}
\label{final_results}
\end{table}
\end{landscape}

\twocolumn

\bibliographystyle{mn2e} 

\bibliography{tf}

\label{lastpage}

\end{document}